\documentclass[aps,prb,reprint,unsortedaddress,superscriptaddress,showpacs]{revtex4-1}
\usepackage{amsmath,amssymb,graphicx,color,bm}
\begin{document}


\title{Theory of tunneling spectroscopy for chiral topological superconductors}


\author{Akihiro Ii}
\affiliation{Department of Applied Physics, Nagoya University, Nagoya 464-8603, Japan}

\author{Ai Yamakage}
\affiliation{Department of Applied Physics, Nagoya University, Nagoya 464-8603, Japan}

\author{Keiji Yada}
\affiliation{Department of Applied Physics, Nagoya University, Nagoya 464-8603, Japan}

\author{Masatoshi Sato}
\affiliation{Institute for Solid State Physics, University of Tokyo, Chiba 277-8581, Japan}

\author{Yukio Tanaka}
\affiliation{Department of Applied Physics, Nagoya University, Nagoya 464-8603, Japan}

\date{\today}

\begin{abstract}
We study the charge conductance of an interface between a normal metal and
a superconducting quantum anomalous Hall system, based on the recursive
 Green's function.
The angle resolved conductance $\gamma(k_{y},eV)$ 
with the momentum $k_{y}$ parallel to the interface and 
the bias voltage $V$ shows a rich structure 
depending on the Chern number ${\cal N}$ of the system. 
We find that when the bias voltage is tuned to the energy dispersion
 of the edge mode, $eV=E_{\rm edge}(k_y)$,
the angle resolved conductance $\gamma(k_y,E_{\rm edge}(k_y))$
shows a pronounced even-odd effect;
the conductance vanishes for  
${\cal N}=0$ or $2$ while it takes a universal value $2e^2/h$
 for ${\cal  N}=1$.
In particular,  in $\mathcal N=2$ phase, we find that the conductance
 $\gamma(k_y,E_{\rm edge}(k_y))$ becomes zero due to interference of
 two degenerate Majorana edge modes, 
although the corresponding surface spectral weight remains non-zero.
\end{abstract}

\pacs{74.45.+c, 74.50.+r, 74.20.Rp}

\maketitle


\section{Introduction}
It is well known  that Andreev bound states (ABSs) are generated
at the edge of unconventional superconductors where the
pair potentials change sign on their
Fermi surfaces. 
\cite{buchholtz81,hara86,kashiwaya00,ABSR2,Hu,TK95,TK96a,ATK04} 
Nowadays, 
the ABSs in unconventional superconductors 
have been recognized as important topological objects in 
condensed matter physics. 
Superconductors hosting 
topologically protected ABSs are dubbed as 
topological superconductor, \cite{Qi,roy06,sato09prb,sato10} and they are
characterized by discrete
symmetries such as the particle-hole symmetry.\cite{schnyder08,ryu10}
Furthermore, it has been clarified as the bulk/edge correspondence that 
when a gapless ABS is generated on the edge,
the corresponding topological invariant exists in the bulk
system.\cite{ryu10,Teo,TanakaReview}

For quasi-two dimensional superconductors, 
according to the energy dispersions,
ABSs are classified into three; flat-type, chiral-type and helical type. 
A flat-type ABS is protected by a one-dimensional winding number which is 
defined for a  fixed $k_{y}$, where $k_{y}$ is the 
momentum parallel to the surface. \cite{Sato11,Yada}
This flat type ABS is 
dubbed as mid gap Andreev bound state and is 
generated for nodal superconductors like
spin-singlet $d_{xy}$-wave one or spin-triplet $p_{x}$-wave one. 
A chiral-type ABS has a linear energy dispersion 
\cite{Matsumoto99,honerkamp98,yamashiro97,Furusaki} and  
is realized in spin-triplet chiral $p$-wave superconductors like 
Sr$_{2}$RuO$_{4}$. \cite{Maeno94,Kashiwaya11}
This ABS is protected by the Chern number\cite{Furusaki,thouless82,
kohmoto85,Qi} and it causes a 
spontaneous current along the surface.  
A chiral ABS has been recognized as a 
chiral Majorana edge mode if the spin degree of freedom is 
quenched.\cite{Qi} \par

By contrast to the case of chiral ABSs, 
the time reversal symmetry is preserved in 
helical ABSs. 
A helical ABS supports two linear energy dispersions with opposite
velocities, which form a Kramers pair.
Instead of a charge current, a spin current is spontaneously generated 
along the edge. 
Stability of the helical ABS is guaranteed by the $\mathbb Z_{2}$ topological
invariant,  
as in the case of quantum spin Hall insulators (QSHs), \cite{Hasan}, and
they are expected to be realized in 
non-centrosymmetric superconductors \cite{bauer04,togano04,nishiyama05,hiller09,reyren07} 
and a bilayer Rashba system,\cite{nakosai12}
where the spin-orbit coupling is important. 
Several new features of  helical ABSs
have been predicted. 
\cite{sato06,sato09,Tanaka,iniotakis07,Eschrig,Eschrig2,Yip,
yokoyama05,iniotakis07,Schnyder2,Schnyder3,Brydon2} 
Surface ABSs in three dimensional systems have been also studied. 
A cone-type ABS is predicted for
a superconducting analogue of the superfluid $^{3}$He B phase. 
This ABS is interpreted as 
Majorana fermion
\cite{Qi,Nagato,Volovik1,Volovik2,Tsutsumi}
obeying massless two dimensional Dirac equation. 
Moreover, surface ABSs with various complicated energy
dispersions\cite{hao11} appear
in superconducting topological insulators, \cite{fu10} e.g.,
Cu$_x$Bi$_2$Se$_3$\cite{hor10,sasaki11}.

A new direction for realization of Majorana fermions 
is to fabricate topological superconductors with conventional pairing.\cite{sato03} 
Especially, much attention has been paid to realize
chiral Majorana modes from the 
view point of 
topological quantum computing. \cite{kitaev03, freedman03, nayak08} 
There are several proposals to fabricate 
Majorana fermions in systems coupled to 
superconductor via the proximity effect. 
It has been proposed that a chiral Majorana edge mode is produced 
at the interface of  ferromagnet/spin-singlet $s$-wave superconductor 
junction on the substrate of three-dimensional topological insulator.
\cite{fu08,fu09,akhmerov09,LLN09,tanaka09,Linder00}
%
Also, a simpler scheme using the Rashba spin-orbit interaction and the
Zeeman field has been proposed. \cite{sato09prl,sato10b,sau10,alicea,LSD10,LSD11,ORO10}
The essential point is the simultaneous presence 
of the strong spin-orbit coupling 
and the time reversal symmetry breaking by the Zeeman field. 
There is another way to realize chiral 
Majorana edge modes by using chiral edge states of a
quantum anomalous Hall system (QAH). \cite{Qi2}
A QAH can be realized by doping of magnetic impurity in 
a QHS. \cite{liu08}
In this scheme, the presence of the 
chiral Majorana edge modes can be controlled by 
the band mass $m$, chemical potential $\mu$ and the pair potential
$\Delta$. 
The number of chiral Majorana edge modes can be 
classified by the Chern number ${\cal N}$ of the system. \cite{Qi}
\par

Stimulated by the idea of 
Qi et al,\cite{Qi2} in our previous paper,  
we have calculated the energy spectrum of the edge states 
and the resulting surface local density of states (SLDOS) for various 
values of the Chern number ${\cal N}$ 
in a heterostructure of 
a QAH and a spin-singlet $s$-wave 
superconductor (QAH$+s$).\cite{ii11} 
To clarify the difference between the 
${\cal N}=1$ and ${\cal N}=2$ states, we applied Zeeman magnetic fields.
We have found  that when the direction of the 
magnetic field is parallel to the interface, the degeneracy 
of the two chiral Majorana edge modes in ${\cal N}=2$ states 
is lifted.  
We have also clarified that the degeneracy is lifted by shifting the chemical 
potential from zero. 
Although the SLDOS has been calculated in detail, 
the relevance to the actual tunneling conductance observed in 
QAH$+s$ system have not been clarified yet.

The purpose of this work is to present a theory of the tunneling
conductance in this system. 
If the ABS has a flat dispersion, which is realized in high-$T_{\rm c}$
cuprate, the tunneling conductance is expressed by
the SLDOS. \cite{Kashiwaya96,TK96}
In the present case, however, the correspondence is not clear.
Since the ABS has a linear dispersion, 
the SLDOS does not always coincide with
the tunneling conductance in normal metal (N)/superconductor junction 
even in the low transparent limit. \cite{Sigrist,YTK97,Matsumoto,Eschrig,Eschrig2}
A similar situation occurs in three dimensions.
Differently from the case of the superconducting analogue of $^3$He B phase, 
\cite{asano03} 
the tunneling conductance for the junction of
N/superconducting topological insulator shows a single zero-bias peak 
by taking into account 
a finite temperature effect\cite{hsieh12} or 
transmissivity at the interface, 
even though the SLDOS has a double peak structure.\cite{yamakage12} 
Because of the difficulty to predict the charge transport property 
from the SLDOS, as mentioned above,
we have to calculate the tunneling conductance of N/(QAH$+s$)/N junction by 
explicitly solving the Bogoliubov-de Gennes (BdG) equation. 

\par
The organization of the paper is as follows. 
In Sec. \ref{sec:2}, 
we  review the model of QAH with spin-singlet $s$-wave 
superconductor. 
In addition, we formulate 
the tunneling conductance in N/(QAH$+s$)/N junction using the recursive
Green's function. 
In Sec. \ref{sec:3a}, we calculate 
the energy dispersion, the SLDOS and the tunneling conductances in
N/QAH/N junction and N/(QAH$+s$)/N one. 
We reveal that an 
even-odd effect in the angle-resolved conductance occurs due to
interference of Majorana fermions whereas the corresponding SLDOS does not.
In  Sec. \ref{sec:4}, we summarize our results. 
\section{Formulation}
\label{sec:2}

In this section we show the model Hamiltonian of QAH$+s$ and the method
of numerical calculation for the SLDOS. 
The model of N/(QAH$+s$)/N junction and the formula of the tunneling
conductance with the recursive Green's function are also shown.

\subsection{Hamiltonian of QAH$+s$}

We consider a QAH on the two-dimensional 
square lattice, which is obtained by the replacement 
$k_{x,y} \rightarrow \sin k_{x,y} $ and $k^2_x+k^2_y \rightarrow
4-2(\cos k_x+\cos k_y)$ in the model used in Refs. \cite{Qi2,chung11}
Near the $\Gamma$ point, this replacement does not change 
the low energy and low wavelength physics of the system. 
Compared to the continuum model, the square lattice model is 
convenient when we calculate the SLDOS.
In the momentum space, the Hamiltonian has the form as $\mathcal H_{\rm QAH}(\bm k) = \bm d(\bm k) \cdot \bm s$ with
\begin{align}
\bm d(\bm k) = (A \sin k_x, A \sin k_y, m(\bm k)),
\end{align}
where $s_i$ is Pauli matrix in spin space and $m(\bm k) = m + 2B(2-\cos
k_x - \cos k_y)$.
The band mass term $m({\bm k})$ determines
the magnitude of the energy shift between up and down spins. 
$A$, $B$ and $m$ are material parameters corresponding to the velocity
of the surface Dirac fermion, the inverse effective mass of
conduction/valence bands, and the band gap, respectively. 
The sign of $m/B$ determines the topological property of the system.
Here note that the presence of $B$ term is crucial to exhibit a QAH. 
The energy dispersion of the above Hamiltonian is symmetric with respect
to the mass term $m$ for $B=0$, but is asymmetric for $B$$\neq$0.  
In other words, a nonzero value of $B$ makes the sign of $m$ meaningful.
Hereafter, we take $A=B=1$ and the lattice constant being unity in our
calculations.

In the following,  we consider the proximity effect by  
an attached spin-singlet $s$-wave superconductor, 
where the pair potential is induced in the QAH (hereinafter we refer to
it as
QAH$+s$).  
The system is described by the BdG Hamiltonian,
\begin{equation}
	{\cal H}_{\rm BdG}(\bm k) = 
	d_z(\bm k)  s_z + [d_x(\bm k) s_x + d_y(\bm k) s_y] \tau_z
	 - \mu  \tau_z + \Delta \tau_x,
\label{eq:BdG}
\end{equation}
where 
$\tau_i$ is Pauli matrix in Nambu space,
$\mu$ is the chemical potential, and $\Delta$ is the induced pair potential 
of spin-singlet $s$-wave superconductor. 
The energy gap of $\mathcal H_{\rm BdG}$ at $\bm k = \bm 0$ is given by $E_{\rm g} = |m| - \sqrt{\Delta^2 + \mu^2}$.
The present system has three phases, i.e. $\mathcal N=0,1$, and
2 phases,\cite{Qi2} which are realized in $m > \sqrt{\Delta^2+\mu^2}$,
$|m| < \sqrt{\Delta^2+\mu^2}$, and $m < -\sqrt{\Delta^2+\mu^2}$,
respectively. 

\subsection{Surface local density of states}

In order to obtain the SLDOS at the 
edge ($x=1$), we introduce an infinite potential barrier at $x=0$.
We calculate the Green's function at $x=1$ 
by $t$-matrix formulation.\cite{Matsumoto1}  
The system is infinite along the $y$-direction 
while it is semi-infinite along the $x$-direction.
Since translational invariance is absent along the $x$-direction,
only the momentum $k_y$ in the $y$-direction is a good quantum number. 
We express the Green's function in the spatial coordinates
$x$ and $x'$ for fixed $k_y$ as follows; 
\begin{align}
	G_{xx'}(k_y,\omega) 
&= {g}_{xx'}(k_y,\omega)
\nonumber \\ & \hspace{2em}
	 -{g}_{x0}(k_y,\omega) {{g}_{00}^{-1}(k_y,\omega)} {g}_{0x'}(k_y,\omega),
\label{eq:green1}
\end{align}
with
\begin{equation}
	{g}_{xx'}(k_y,\omega) = \frac{1}{N_x} \sum_{k_x} e^{ik_x(x-x')} {g}(k_x,k_y,\omega),
\end{equation}
and 
\begin{equation}
	{g}^{-1}(k_x,k_y,\omega) = {\omega-{\mathcal H_{\rm BdG}}(k_x,k_y)},
\end{equation}
where $N_x$ is the total number of 
lattice points in the $x$-direction. In the right hand side of
Eq.(\ref{eq:green1}), 
the first term denotes the unperturbed bulk 
Green function, and the second term comes from
the scattering effect at the edge. 
The angle resolved 
SLDOS $N(k_{y},\omega)$ at $x=1$  is written as
\begin{align}
N(k_y,\omega) &= -\frac{1}{\pi} {\rm Im} \mathrm{Tr} \left[P_{\rm e} {G}_{11}^{\rm R}(k_y,\omega) \right],
\end{align}
where 
\begin{align}
G^{\rm R}_{xx'}(k_y, \omega) = G_{xx'}(k_y,\omega+ i\eta),
\label{retarded}
\end{align}
 is the retarded Green's function, $\eta$ is an infinitesimal positive number,
and $P_{\rm e} = (1+\tau_z)/2$ is the projection operator onto the particle subspace.
From the above equations, one obtains the SLDOS $D(\omega)$ for energy $\omega$ measured from 
the Fermi level as follows
\begin{equation}
	D(\omega) = \frac{1}{N_y} \sum_{k_y} N(k_y,\omega),
\end{equation}
where $N_y$ is a total number of lattice points for the $y$-direction. 
In the actual calculation, we set $N_x=N_y=4096$.

\subsection{N/(QAH$+s$)/N junction}

\begin{figure}
\begin{center}
	\includegraphics[width=8.0cm]{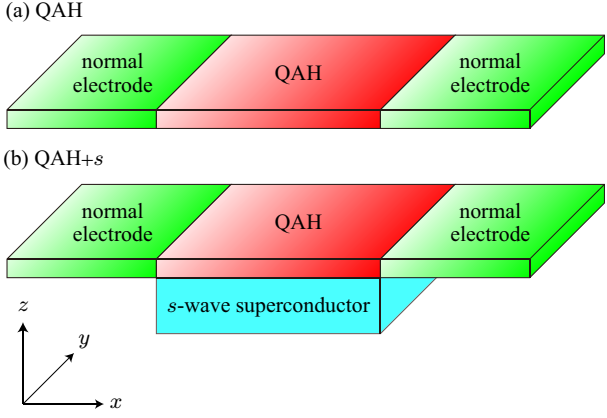}
	\caption{(color online) Schematic illustrations 
of N/QAH/N (a) and N/(QAH$+s$)/N (b) junctions .}
	\label{fig.1}
\end{center}
\end{figure}

Now we show the
Hamiltonian of N/(QAH$+s$)/N junction illustrated in Fig.\ref{fig.1}. 
The Hamiltonian of QAH$+s$ in the center region of the junction
is represented as 
\begin{align}
 \mathcal H_{\rm BdG}(k_y)
 &= \sum_{x=1}^{N_{\mathrm{QAH}}} c^\dag_x(k_y) \epsilon(k_y) c_x(k_y)
 \nonumber \\ &  +
 \sum_{x=1}^{N_\mathrm{QAH}-1} \left[ c^\dag_x(k_y) t_{\rm QAH} c_{x+1}(k_y) + \mathrm{h.c.} \right],
\end{align}
with 
\begin{align}
 \epsilon(k_y) = [m+2B(2-\cos k_y)] s_z + A \sin k_y s_y  \tau_z + \Delta \tau_x,
 \end{align}
 and
 \begin{align}
 t_{\rm QAH} = -B s_z -i A s_x \tau_z / 2. 
\end{align}

The Hamiltonians of normal electrodes located in the left ($\mathcal H_{\rm L}$) and the right ($\mathcal H_{\rm R}$) regions are given by
\begin{align}
 \mathcal H_{\rm L}(k_y) &= 
\sum_{x=-\infty}^0 c^\dag_x(k_y) 
(-2t_{\rm N} \cos k_y - \mu_{\rm N}) \tau_z c_x(k_y)  
\nonumber\\ & \hspace{-1em} +
\left(  \sum_{x=-\infty}^{-1} c^\dag_x(k_y) (-t_{\rm N}) \tau_z
 c_{x+1}(k_y) + \mathrm{h.c.} \right), 
\end{align}
\begin{align}
 \mathcal H_{\rm R}(k_y) &= 
\sum_{x=N_{\rm QAH}+1}^\infty c^\dag_x(k_y) (-2t_{\rm N} \cos k_y 
- \mu_{\rm N}) \tau_z c_x(k_y)  
\nonumber\\ & \hspace{-3em} +
\left( \sum_{x=N_{\rm QAH}+1}^{\infty} 
c^\dag_x(k_y) (-t_{\rm N}) \tau_z c_{x+1}(k_y) + \mathrm{h.c.} \right).
\end{align}
We also assume the following simple hopping $\mathcal H_{\rm j}$ between the
electrode and the QAH$+s$,
\begin{align}
 \mathcal H_{\rm j}(k_y) 
&= c^\dag_{0}(k_y) (-t_{\rm j}) \tau_z c_{1}(k_y)
 \nonumber\\ & \quad 
 +
 c^\dag_{N_{\rm QAH}}(k_y) (-t_{\rm j}) \tau_z c_{N_\mathrm{QAH}+1}(k_y)
 + \mathrm{h.c.}
\end{align}
In the actual calculation, 
$t_{\rm j}$ is fixed as $t_{\rm j} = t_{\rm N}$, 
for simplicity.

\subsection{Tunneling conductance and \\recursive Green's function}

The angle resolved 
tunneling conductance $\gamma(p_y,\omega)$ in a junction is given by the Lee-Fisher formula\cite{lee81}
\begin{align}
 \gamma(p_y,\omega) &= \frac{t_{\rm j}^2e^2}{2h}
 \mathrm{Tr} \bigl[P_{\rm e} (
  G_{x,x+1}'' G_{x,x+1}'' + G_{x+1,x}''G_{x+1,x}'' 
\nonumber \\ & \hspace{2em}
- G_{x,x}''G_{x+1,x+1}'' - G_{x+1,x+1}''G_{x,x}''
 )\bigr],
 \label{gamma}
\end{align}
with $G''_{xx'} = \mathrm{Im} \, G^{\rm R}_{xx'}$.
Due to current conservation in the normal metals, 
we can choose arbitrary $x$ for Eq.(\ref{gamma}) in $x < 0$ or $x \geq N_{\rm QAH}+1$, except in the superconducting region ($1 \leq x \leq N_{\rm QAH}$).
The total conductance $\Gamma$ is given by $\Gamma(\omega) = \sum_{p_y} \gamma(p_y,\omega)$.
We first calculate the Green's function $G_{\mathrm L,l,l}$ in the left semi-infinite system where the sites in $x>l$ are deleted.
$G_{\mathrm L, x,x}$ satisfies the following recursive relation:
\begin{align}
 G_{\mathrm L, x, x}^{-1} = g_x^{-1} - \mathcal H_{x,x-1} G_{\mathrm L, x-1,x-1} \mathcal H_{x-1,x},
 \label{GL}
\end{align}
with $\mathcal H_{x,x'}$ being the hopping from $x'$ to $x$.
Here,
$g_x^{-1}(p_y,\omega) = \omega - \mathcal H_{x,x}(p_y)$ is the Green's function in the isolated $x$-th column.
In the present model, 
only $\mathcal H_{x,x'}$  with $|x-x'| \leq 1$ is nonzero, and  given by
\begin{align}
\mathcal H_{x,x'} &= [(-2t_{\rm N} \cos k_y - \mu_{\rm N})\delta_{x,x'}  -t_{\rm N}\delta_{x',x\pm1}] \tau_z,
\end{align}
for $x, x' \leq 0$ or $x, x' \geq N_{\rm QAH}+1$, and
\begin{align}
\mathcal H_{x,x'} &= \epsilon(k_y) \delta_{x,x'} + t_{\rm QAH} \delta_{x',x+1} + t_{\rm QAH}^\dag \delta_{x',x-1},
\end{align}
for $1 \leq x,x'  \leq N_{\rm QAH}$.
At the interfaces, $\mathcal H_{0,1}$, $\mathcal H_{1,0}$, $\mathcal H_{N_{\rm QAH},N_{\rm QAH}+1}$ and $\mathcal H_{N_{\rm QAH}+1, N_{\rm QAH}}$ are given by
\begin{align}
 \mathcal H_{0,1} 
&=\mathcal H^\dag_{1,0} 
= 
\mathcal H_{N_{\rm QAH}, N_{\rm QAH}+1}^\dag
=\mathcal H_{N_{\rm QAH}+1, N_{\rm QAH}} 
\nonumber \\
& = -t_{\rm j} \tau_z.
\end{align}
The Green's function in the right semi-infinite system $G_{\mathrm R, l, l}$ where the sites in $x<l$ are deleted satisfies the following relation.
\begin{align}
 G_{\mathrm R, x, x}^{-1} = g_x^{-1} - \mathcal H_{x,x+1} G_{\mathrm R, x+1,x+1} \mathcal H_{x+1,x}.
 \label{GR}
\end{align}
It is noted that the Green's functions at the edge of the electrode ($G_{\rm L,0,0}$ and $G_{\rm R, N_{\rm QAH}+1,N_{\rm QAH}+1}$) are obtained by Eq.(\ref{eq:green1}).
Then, using Eqs. (\ref{GL}) and (\ref{GR}), we can recursively obtain $G_{\mathrm L,x,x}$ and $G_{\mathrm R, x, x}$ for any $x$.
The site-diagonal part of the Green's function is obtained in terms of
the above Green's functions,
\begin{align}
 G_{x,x}^{-1}
&= g_x^{-1} - \mathcal H_{x,x-1} G_{\mathrm L, x-1,x-1} \mathcal H_{x-1,x} 
\nonumber \\ & \hspace{4em}
- \mathcal H_{x,x+1} G_{\mathrm R, x+1,x+1} \mathcal H_{x+1,x},
\end{align}
and the site-off-diagonal parts are also obtained as 
\begin{align}
G_{x,x+1} &= G_{x,x} \mathcal H_{x,x+1} G_{\mathrm R, x+1,x+1},
\\
G_{x+1,x} &= G_{x+1,x+1} \mathcal H_{x+1,x} G_{\mathrm L, x,x}.
\end{align}
We can calculate the conductance $\Gamma$ from $G_{x,x}$, $G_{x+1,x+1}$,
$G_{x,x+1}$, and $G_{x+1,1}$ by using Eq. (\ref{gamma}).

\section{Results and discussions}
\label{sec:3a}

In this section, we show our numerical results for the electronic states and
the tunneling conductance of the QAH$+s$.
Experimental proposals to detect our results are also discussed.

\subsection{Electronic states and tunneling conductance in N/QAH/N junction}
\begin{figure}
\includegraphics[scale=0.765]{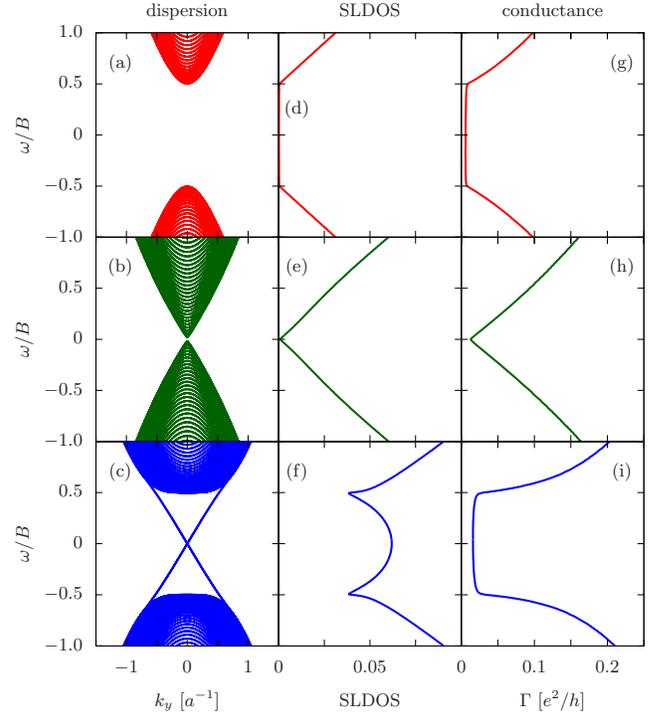}
	\caption{(color online) Energy dispersion relations [(a),(b), and (c)], SLDOS [(d), (e), and (f)], and the tunneling conductances in N/QAH/N junction, 
for the case with the trivial insulator [(g):$m/B=0.5$], the critical point [(h):$m=0$], and the QAH [(c):$m/B=-0.5$]. 
$a$ denotes the lattice constant.
The parameters are taken as follows.
$A/B=1$, $\Delta=\mu=0$, and $N_{\rm QAH} = 100$ for the energy dispersion, $N_{\rm QAH}=4096$ for the SLDOS, $N_{\rm QAH}=1000$ for the conductance.
}
	\label{fig.2}
\end{figure}

Before discussing the superconducting case, we check the electronic
states and  the tunneling conductance of the N/QAH/N junction at $\Delta=
\mu = 0$.
The energy dispersions with the finite width ($N_{\rm QAH}=100$) are
shown in Figs. \ref{fig.2}(a), (b), and (c).
There is a band gap in  
Figs. \ref{fig.2}(a) and \ref{fig.2}(c), respectively,  
while it closes at the critical point with $m=0$ [Fig. \ref{fig.2}(b)].
Gapless chiral edge modes show up when $m < 0$ 
[Fig. \ref{fig.2}(c)].
Note that two gapless modes propagating in opposite directions appear in
Fig. \ref{fig.2}(c) since both the left ($x=0$) and right edges
($x=N_{\rm QAH}$) are present in the calculation, i.e.,
each edge has an edge state.
Figures \ref{fig.2}(d), (e), and (f) show the SLDOS at the edge.
In the trivial insulator phase with $m \geq 0$ [Fig. \ref{fig.2}(d) and (e)],  
the line shapes of the SLDOS are the same as those in the bulk. 
On the other hand, in the case of QAH, the SLDOS is enhanced in the band gap due
to the gapless edge modes, as shown in Fig. \ref{fig.2}(f).

The line shapes of tunneling conductance  are similar to those of the SLDOS
when $m \geq 0$ [Fig.\ref{fig.2}(g) and (h)], i.e., U-shaped
gap in $|\omega| < m$ [Fig.\ref{fig.2}(g)] and V-shaped dip at $m=0$
[Fig.\ref{fig.2}(h)].
In the case of QAH, although the SLDOS shows a zero-energy
peak [Fig.\ref{fig.2}(f)], the corresponding tunneling conductance shows
a U-shaped gap [Fig.\ref{fig.2}(i)] similar to that in the case with
$m>0$ [Fig. \ref{fig.2}(g)].
This is because the central region of
the junction has a bulk gap, thus the tunneling conductance should be zero. 

\begin{figure}
\includegraphics[scale=0.8]{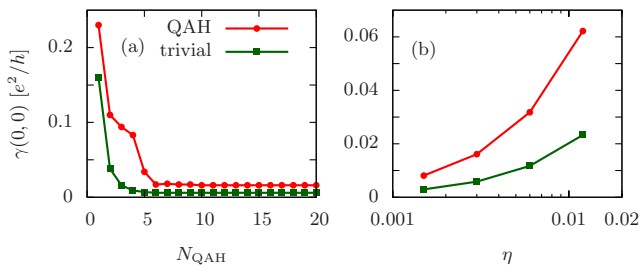}
	\caption{
	(color online)
Angle resolved tunneling conductance $\gamma(k_y,\omega)$ at $k_y=0$ and $\omega=0$ as functions of $N_{\rm QAH}$ (a) and  $\eta$ (b), for the QAH with $m/B=-0.5$ and the trivial insulator with $m/B=0.5$ and $N_{\rm QAH} = 1000$. 
The parameters are the same as in Fig. \ref{fig.2}.}
	\label{fig.3}
\end{figure}

Here we notice that the conductance in Figs. \ref{fig.2}(g) and
(i) takes a small but non-zero value in the energy gap, but this comes
from a nonzero value of $\eta$ in our numerical calculation.
Indeed, 
as one decreases $\eta$ in Eq. (\ref{retarded})  and increases the
number of QAH layer ($N_{\rm QAH}$), 
$\gamma(0,0)$ tends to be zero, as shown
in Fig. \ref{fig.3}, i.e., the corresponding conductance goes to zero in
the limit of $N_{\rm QAH} \to \infty$ and
$\eta \to +0$. 
Additionally, we note that the tunneling conductance of the QAH is lager
than that of the trivial insulator for finite $N_{\rm QAH}$ and 
$\eta$ (Fig.\ref{fig.3}). 
This is due to hybridization of the edge states located at $x=0$ and $x=N_{\rm QAH}$. 

\subsection{Electronic states of QAH$+s$}

\begin{figure}
\begin{center}
\includegraphics[scale=0.8]{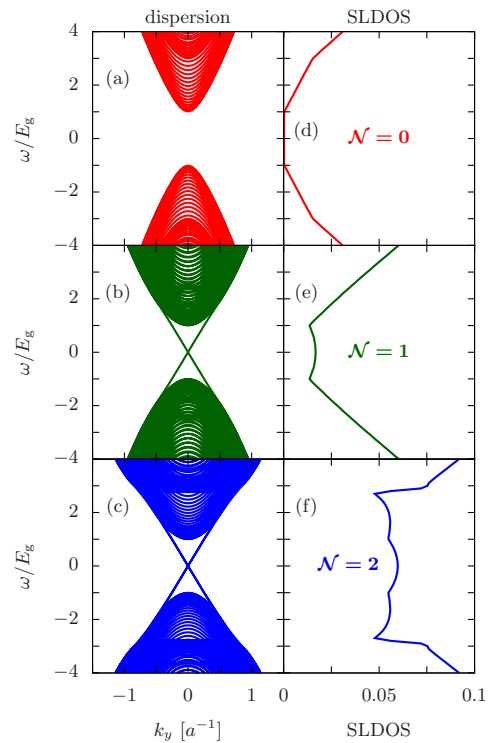}
	\caption{(color online) Energy dispersion relations and the SLDOSs in QAH$+s$ with $\mathcal N=0$ [(a) and (d): $m/B=0.5$], $\mathcal N=1$ [(b) and (e): $m=0$], and $\mathcal N=2$ [(c) and (f): $m/B=-0.5$].
The pair potential is taken to be $\Delta/B=0.25$, and the other parameters are the same as in Fig. \ref{fig.2}.
}
	\label{fig.4}
\end{center}
\end{figure}

Now we consider the superconducting case with $\Delta/B = 0.25$.
The trivial superconductor with $\mathcal N=0$ does not have any gapless
state as shown in Fig. \ref{fig.4}(a), while the topological
superconductors with $\mathcal N=1$ [Fig.\ref{fig.4}(b)] and $\mathcal
N=2$ [Fig.\ref{fig.4}(c)] have.
Although the gapless modes in the latter two phases have a similar
energy dispersion, we can 
distinguish them by the SLDOS, as shown in Figs. \ref{fig.4}(e) and (f):
The line shape of the SLDOS for $\mathcal N=1$ shows a zero-bias peak
[Fig.\ref{fig.4}(e)]. 
On the other hand, that for $\mathcal N=2$ shows a larger zero-bias peak
and satellite peaks at $\omega/E_{\rm g} \sim \pm 2$
[Fig. \ref{fig.4}(f)].
The larger zero-bias peak is due to two gapless modes, and the
satellite peaks come from a branch of ABS near the  
bulk bands.

\subsection{Even-odd effect in the angle resolved conductance}

\begin{figure}
\includegraphics[scale=0.65]{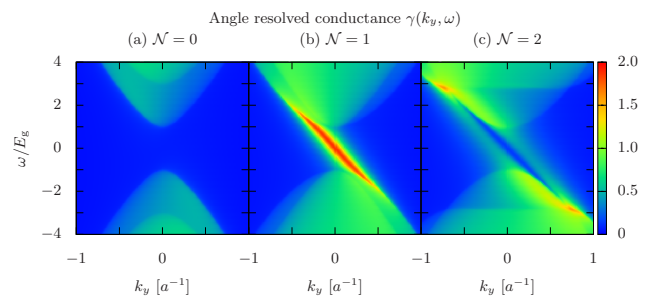}
	\caption{(color online) Angle resolved conductance in N/(QAH$+s$)/N junction with $\mathcal N=0$ (a), $\mathcal N=1$ (b), and $\mathcal N=2$ (c) phases.}
	\label{fig.5}
\end{figure}

Next we discuss the angle resolved conductance $\gamma(k_y,\omega)$ in
the N/(QAH$+s$)/N junction shown in Fig. \ref{fig.5}.
In $\mathcal N=0$ [Fig. \ref{fig.5}(a)] and $\mathcal N=1$
[Fig. \ref{fig.5}(b)] phases, the conductance spectra are naturally
understood by the energy spectra [Figs. \ref{fig.4}(a)(b)] and the
SLDOSs [Figs. \ref{fig.4}(d)(e)].
Due to resonance between the incident state and the chiral edge mode,
the value of conductance in $\mathcal N=1$ phase takes $\gamma(k_y,
\omega) \sim 2 e^2/h$ at $\omega = E_{\rm edge}(k_y)$, where $E_{\rm
edge}(k_y)$ is the energy dispersion relation of the edge state. 
Note that the conductance spectra are asymmetric with respect to $k_y=0$
because the present edge mode is \textit{chiral}.

In $\mathcal N=2$ phase, we obtain a remarkable result: In this phase,
it is natively expected that the tunneling conductance take a doubled
value of that in $\mathcal N=1$ phase since there are two edge modes.
It is, however, not the case.
As shown in Fig. \ref{fig.5}(c),
in $\mathcal N=2$ phase, the conductance takes a {\it smaller} value
than that in $\mathcal N=1$ phase at
$\omega \sim E_{\rm edge}(k_y)$.
In particular, the conductance vanishes just at $\omega = E_{\rm edge}(k_y)$.
\begin{figure}
\begin{center}
\includegraphics[scale=0.79]{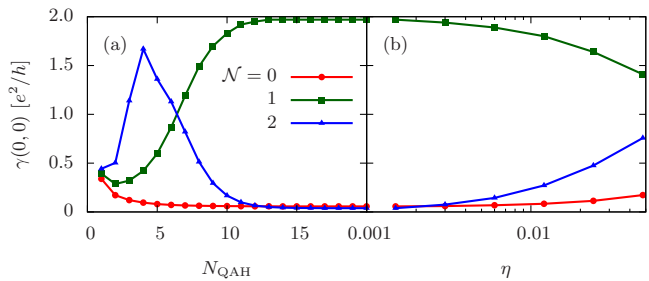}
	\caption{(color online) Angle resolved conductance $\gamma$ at
 $k_y=0$ and $\omega=0$ as functions of $N_{\rm QAH}$ (a) and $\eta$
 (b).} 
	\label{fig.6}
\end{center}
\end{figure}
To confirm this, we study $\gamma(k_y,\omega)$ at $k_y=0$ and $\omega=0$
as functions of $N_{\rm QAH}$ [Fig. \ref{fig.6}(a)] and $\eta$
[Fig. \ref{fig.6}(b)]. 
It is found that as $N_{\rm QAH}$ increases and $\eta$ decreases,
$\gamma(0,0)$ converges to $2e^2/h$ in $\mathcal N=1$ phase, and to 0 in
$\mathcal N=0,2$ phases. 
Therefore, the conductance shows an \textit{even-odd effect} as $\gamma =
[1-(-1)^{\mathcal N}] e^2/h$ in the presence of chiral Majorana
fermions.

The vanishment of the tunneling conductance 
originates from the degeneracy of Majorana edge fermions in ${\cal N}=2$
phase: Indeed, it
is suppressed by lifting the degeneracy by tuning the chemical
potential or applying Zeeman fields. 
At a finite chemical potential, the degeneracy of
the gapless modes is lifted and 
\begin{figure}
\includegraphics[scale=0.75]{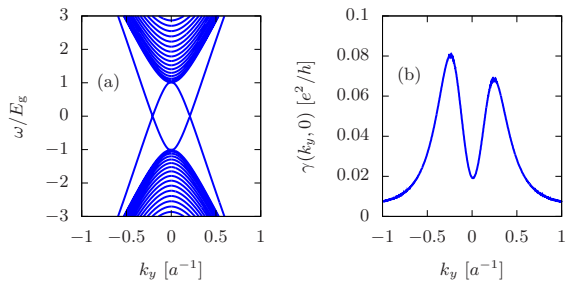}
	\caption{(color online) The energy dispersion of the QAH$+s$ (a) and the angle resolved conductances (b) in $\mathcal N=2$ phase.
The chemical potentials are set to be $\mu/B=0.3$.
The energy gap is given by $E_{\rm g}/B = 0.11$.
}
	\label{fig.8}
\end{figure}
the zero energy states appear at the two finite $k_y$ points, as shown
in Fig. \ref{fig.8}(a). 
The conductance at $\omega=0$ with
the finite $\mu$ is shown in Fig. \ref{fig.8}(b).
Contrary to that with
$\mu=0$, $\gamma(0,0)$ remains non-zero.
This result suggests that the vanished conductance arises from
interference of degenerated Majorana fermions.

Here we note that the degeneracy of Majorana edge fermions is ensured by
a symmetry of the system: 
When $\mu=0$, the BdG Hamiltonian (\ref{eq:BdG}) has the following
additional symmetry, 
\begin{eqnarray}
s_y\tau_z {\cal H}_{\rm BdG}(k_x,k_y)s_y\tau_z=-{\cal H}_{\rm BdG}(k_x,-k_y), 
\end{eqnarray}
and on the $k_x$ axis, this reduces to the so called chiral symmetry,
\begin{eqnarray}
\{\Gamma, {\cal H}_{\rm BdG}(k_x, 0)\}=0, 
\end{eqnarray}
with $\Gamma=s_y\tau_z$.
Thus, following Refs.\onlinecite{sato09, Sato11}, one can introduce the
one-dimensional winding number,
\begin{eqnarray}
W=-\frac{1}{4\pi i}\int_{-\pi}^{\pi}dk_x 
{\rm tr}\left[\Gamma {\cal H}_{\rm BdG}^{-1}\partial_{k_x}{\cal H}_{\rm
	 BdG}\right]_{k_y=0},  
\end{eqnarray}
which can be evaluated as $W=2$ in the case of ${\cal N}=2$.\cite{ii11}.
Therefore, the bulk-edge correspondence ensures that there exist
two degenerate Majorana edge modes at $k_y=0$. 
As we mentioned above, because the vanishment of the tunneling conductance
occurs only when Majorana edge modes are degenerate, it is very likely
that this chiral symmetry is responsible for the destructive
interference of the tunneling conductance reported here.  
 

In the normal (not superconducting) states,
even-odd effects in conductance appear in mono/bi-layer graphene,
\cite{katsnelson06} which can be generalized to the system with
spin-orbit interactions. \cite{yamakage09}
They are interpreted as a result of mirror symmetry of the
system.\cite{yamakage11}
Also, graphene nano-ribbons show even-odd effects in the conductance
\cite{akhmerov08,cresti08,nakabayashi09}, which can be understood using
parity of the system.
We believe that our result is also explained in the viewpoint of symmetry.

On the other hand, in superconducting states,
various even-odd effects of Majorana fermions have been reported so far. 
For instance,  
in N / a chain of Majorana bound states junction, 
the tunneling conductance shows an even-odd effect as a function of the
length of the chain. \cite{Flensberg10} 
It has been also known that the SLDOS at the zero energy in a multiband
Rashba superconductor with Zeeman interaction
shows an even-odd effect as a function of the number of occupied
subband.\cite{potter11} 
In these cases, no degeneracy of Majorana fermion exists by
hybridization when the number of the Majorana fermions is even. 
On the other hand, in our case, the degeneracy in ${\cal N}=2$ phase is
essential to obtain the even-odd effects.
Therefore, the even-odd effect reported in the present paper is
essentially distinct from the previous ones, and it originates from the
interference without using interferometers as discussed in
Refs. \onlinecite{fu09,akhmerov09,sau11}

\subsection{Proposals for experiment}

%
%

Before closing the section, 
we propose how to detect the even-odd effect mentioned above.
The simplest observable is the angle-integrated tunneling conductance.
\begin{figure}
\includegraphics[scale=1]{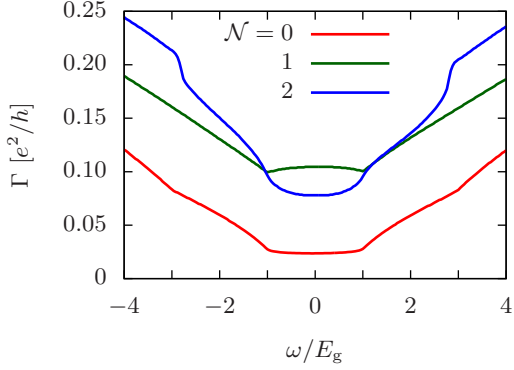}
	\caption{(color online) Tunneling conductances in the N/(QAH$+s$)/N junction with $\mathcal N=0$, $\mathcal N=1$, and $\mathcal N=2$ phases.}
	\label{fig.9}
\end{figure}
In $\mathcal N=0$ phase, the system has no gapless mode, then the value
of conductance becomes smaller in the energy gap, as shown in
Fig. \ref{fig.9} (red line).
In $\mathcal N=1$ phase, the line shape of conductance shows a zero-bias
peak due to gapless mode [Fig. \ref{fig.9} (green line)]. 
On the other hand,
in $\mathcal N=2$ phase, 
the line shape of conductance shows a zero-bias dip [Fig. \ref{fig.9}
(blue line)], in spite of the presence of gapless modes. 

The more direct evidence is to observe the angle resolved current by
scanning a charged tip above the system as was performed for
two-dimensional electron gases in GaAs
heterostructures. \cite{topinka00,topinka01} 
Moreover, 
it is useful  to fabricate the nanoribbon of N/(QAH$+s$)/N junction
since  normal incident electron with $k_y=0$ effectively contributes to
$\gamma(0,0)$.


%
%
%
%
%
%
\section{Summary}
\label{sec:4}

We studied the tunneling conductances of N/(QAH$+s$)/N junction in
$\mathcal N=0$, $\mathcal N=1$, and $\mathcal N=2$ phases.
In the presence of gapless edge modes in $\mathcal N=1$ and $\mathcal
N=2$ phases, the corresponding SLDOSs take finite values in the bulk
energy gap.
In $\mathcal N=1$ phase  the angle resolved conductance also takes
the finite value of $\gamma(k_y, \omega) = 2e^2/h$ when the incident
electron is resonant with the chiral edge mode  at $\omega = E_{\rm
edge}(k_y)$.
On the contrary, in $\mathcal N=2$ phase, the tunneling conductance
vanishes at $\omega = E_{\rm edge}(k_y)$ although the corresponding
SLDOS does not.
This stems from the interference of the degenerated Majorana fermions.  
Namely, an even-odd effect with respect to the number of Majorana
fermion $\mathcal N$ occurs. 

Although we have partly addressed the mechanism of the even-odd effect, it has been not definitely answered.
To reveal this, 
the following things are needed to be unveiled:
relation between the chiral symmetry and the tunneling conductance, robustness of the even-odd effect against disorder proved by the microscopic calculation, 
and the even-odd effect for the higher Chern number of $\mathcal N \geq 3$.
We will study these issues in the future work.


\begin{acknowledgments}
This work
was supported by MEXT (Innovative Area ``Topological
Quantum Phenomena" KAKENHI), and in part by
the National Science Foundation under Grant No. NSF
PHY05-51164.
\end{acknowledgments}

\appendix
\section{Tunneling conductance of N/(QAH$+s$) junction in the continuum limit}

To confirm the even-odd effect found in this paper, 
we study the tunneling conductance of N/(QAH$+s$) junction in the
continuum limit.

Let us consider a normal metal in the left side ($x<0$), whose
Hamiltonian is given by
\begin{align}
 H_{\rm N}(\bm k) = \left( \frac{k^2}{2m_{\rm e}} -\mu_{\rm N} \right) \tau_z,
\end{align}
and QAH$+s$ in the right side ($x>0$).
Here $k=(k_x^2+k_y^2)^{1/2}$ is the magnitude of the two-dimensional momentum.
The  Hamiltonian of QAH$+s$ is obtained by the $k \cdot p$ theory as \cite{Qi2}
\begin{align}
 H(\bm k) &= m(\bm k) s_z + A a ( k_x s_x + k_y s_y) \tau_z -\mu \tau_z + \Delta \tau_x,
 \\
 m(\bm k) &= m + B (ka)^2.
\end{align}
The eigenvalue of the above Hamiltonian is given by
\begin{align}
 E_{\alpha\beta}(\bm k) &= \alpha \Bigl[m^2(\bm k) + A^2 k^2 a^2 + \mu^2 + \Delta^2 +  
\nonumber\\ & +
2\beta \sqrt{(m^2(\bm k) + A^2 k^2 a^2) \mu^2 + m^2(\bm k)\Delta^2} \Bigr],
\end{align}
where $\alpha, \beta = \pm$. 
The corresponding eigenvector $\bm u_{\alpha\beta}(\bm k)$ is also obtained analytically.

Now we calculate the tunneling conductance,
generalizing theories of the tunneling
spectroscopy of conventional \cite{BTK_B82} and unconventional
\cite{TK95,KT_R00} superconductors. 
The wave function in the normal metal ($x<0$) is given by
\begin{align}
 \psi_{\mathrm N, s}(\bm x) 
&= 
\Bigl[
 \chi_{s\mathrm e} e^{i k_{\mathrm e x} x}
 + \sum_{s'} \bigl(b_{ss'} \chi_{s' \mathrm e} e^{ -ik_{\mathrm e x}x} 
\nonumber \\ & \hspace{7em}
+ a_{ss'} \chi_{s' \mathrm h} e^{i k_{\mathrm h x} x} \bigr) \Bigr] e^{i k_y y},
\end{align}
where $\chi_{s \tau}$ is the eigenvector of $H_{\rm N}({\bm k})$ with spin $s$ for electron $(\tau = \mathrm e)$ or hole $(\tau
= \mathrm h)$, and 
$k_{\mathrm e x} = ({k_{\rm e}^{2} - k_y^2})^{1/2} = k_{\rm e} \cos \theta$, 
$k_{\rm e} = \sqrt{2m_{\rm e}(\mu_{\rm N} + E)}$, 
$k_{\mathrm h x}
= [{2m_{\rm e}(\mu_{\rm N} - E) - k_y^2 }]^{1/2}$. 
The first term of the wave function denotes
an injected electron, and the second (third) one denotes 
a reflected hole (electron)
with reflection coefficient $a_{s s'}$ ($b_{s s'}$). 
The wave function in the QAH$+s$ ($x > 0$) is given
by 
\begin{align}
 \psi_{\mathrm{QAH}+s}(x) = \sum_i t_i \bm u_i e^{i (q_i x + k_y y)},
\end{align}
where $q_i$, $(i = 1,\cdots,4)$ is a solution of $E =
E_{\alpha_i\beta_i}(q_i, k_y)$.
Among the eigenvectors,  $\psi_{\mathrm{QAH}+s}(\bm x)$ consists of those with
$E_{\alpha_i\beta_i}(q_i,k_y)/\partial q_i > 0$ or $\mathrm{Im} (q_i) > 0$, where the former
denotes right-going states and the latter denotes localized
states in the vicinity of $x = 0$.
These wave functions are connected at the interface ($x = 0$)
by the conditions \cite{molenkamp01},
$\psi_{\rm N}(0) =\psi_{\mathrm{QAH}+s}(0)$ and 
$v_{\rm N}\psi_{\rm N}(0)=v_{\mathrm{QAH}+s}\psi_{\mathrm{QAH}+s}(0)$,
with the velocity operator
$v_{\mathrm {N}(\mathrm{QAH}+s)} = \partial H_{\mathrm{N}(\mathrm{QAH}+s)} /
\partial k_x|_{k_x \to -i\partial_x}$.  
The above equations determine the
coefficients $a_{s s'}$, $b_{s s'}$ and $t_{i}$. 
Finally, the charge conductance $\gamma(k_y,\omega)$ is given by
\begin{align}
\gamma(k_y,\omega) = 
\frac{e^2}{h}
\left[ 2 + \sum_{ss'}
\left(|a_{s s'}|^2 - |b_{s s'}|^2 \right) \right] .
\end{align}
In the following, the material parameters of the normal metal are fixed as $m_{\rm e} B a^2 = 1$, $\mu_{\rm N}/B = 100$, and
the material parameters of QAH$+s$ are chosen as $A = B$ and
$\Delta/B = 0.25$.

\begin{figure}
\includegraphics[scale=0.9]{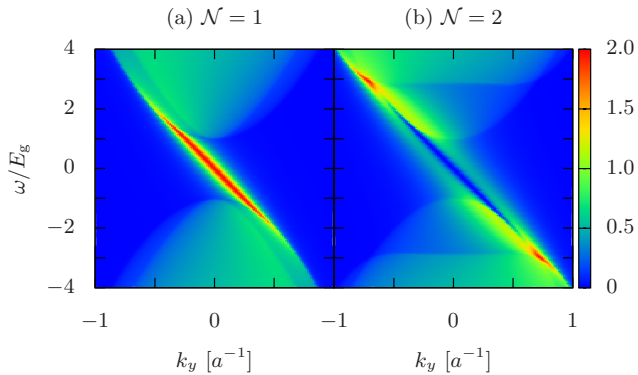}
\caption{(color online) Angle resolved conductance $\gamma(k_y,\omega)$ of the N/(QAH$+s$) junction for $\mathcal N=1$ with $\tilde \mu = \mu/B = 0$ (a) and $\mathcal N=2$ with $\tilde \mu = 0.3$ (b) phases. $E_{\rm g} = 0.25 B$ is the magnitude of the superconducting gap.
}
\label{gamma_ky_omega}
\end{figure}

\begin{figure}
\includegraphics[scale=0.93]{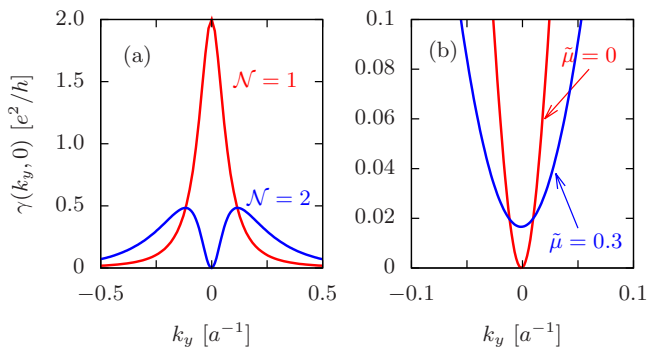}
\caption{(color online) The angle resolved tunneling conductance
 $\gamma(k_y,0)$ at the zero-bias voltage $\omega=0$ for $\mathcal N=1$
 and $\mathcal N=2$ phases (a). 
Those in $\mathcal N=2$ phase for $\tilde \mu = \mu/B = 0$ and $\tilde \mu = 0.3$ are also shown (b).
}
\label{btk}
\end{figure}

The obtained angle resolved tunneling conductances are shown in
Fig. \ref{gamma_ky_omega}.
These spectra are consistent with those obtained in the lattice model
shown in Fig. \ref{fig.5}; the value of $\gamma(k_y,E_{\rm
edge}(k_y))$ takes $2 e^2/h$ in $\mathcal N=1$ phase
[Fig. \ref{gamma_ky_omega}(a)] while it takes 0 in $\mathcal N =2$ phase
[Fig. \ref{gamma_ky_omega}(b)].
In order to see the even-odd effect more clearly,
we focus on $\gamma(k_y,0)$.
Figure \ref{btk}(a) shows the angle resolved conductance at the
zero-bias voltage ($\omega=0$), where the branch of Majorana fermions appears at
$k_y=0$, i.e., $E_{\rm edge}(0)=0$, both for $\mathcal N = 1$ and $\mathcal N
= 2$  phases.
The value of $\gamma(0,0)$ takes $2e^2/h$ in $\mathcal N=1$ phase and
$0$ in $\mathcal N=2$ phase. 
However, if one tunes the chemical potential $\mu$ away from zero, where
the degeneracy of two Majorana fermions is lifted, the value of
conductance recovers to be finite at $k_y=0$, as shown in
Fig. \ref{btk}(b). 

As compared to the calculation in the lattice system, 
the present approach in the appendix has advantages, i.e., it is easy to
take the thermodynamic limit, and the infinitesimal small factor $\eta$
is not necessary. 
The above result indicates that the even-odd effect found in
this paper is robust, and the vanishing conductance is
driven by the degenerating two Majorana fermions.

\newpage
\bibliography{majorana_interference}

\end{document}